\documentstyle[preprint,aps,epsfig]{revtex} 
\begin{document} 
\title{Competition between Pauli and orbital effects in a charge-density wave system} 
\author{J. S. Qualls$^1$, L. Balicas$^{1,2}$, J. S. Brooks $^1$, N. Harrison$^{3}$, L. K. Montgomery $^{4}$, and M. Tokumoto$^{5}$} 
\address{$^1$National High Magnetic Field Laboratory, Florida State University, Tallahassee, FL 32306, USA} 
\address{$^2$Instituto Venezolano de Investigaciones Cient\'{\i}ficas, apartado 21827, Caracas 1020A - Venezuela} 
\address{$^3$National High Magnetic Field Laboratory, LANL, MS-E536, Los Alamos, NM 87545, USA } 
\address{$^4$Department of Chemistry, Indiana University, Bloomington, IN 47405, USA} 
\address{$^5$Electrotechnical Laboratory, Tsukuba, Ibaraki 305, Japan } 
\maketitle 
\date{Received: \today }

\begin{abstract} 
We present angular dependent magneto-transport and magnetization 
measurements on $\alpha$-(ET)$_{2}$MHg(SCN)$_{4}$ compounds at 
high magnetic fields and low temperatures. We find that the low 
temperature ground state undergoes two subsequent
field-induced density-wave type phase transitions above a critical angle of the 
magnetic field with respect to the crystallographic axes. This new 
phase diagram may be qualitatively described assuming a charge 
density wave ground state which undergoes field-induced 
transitions due to the interplay of Pauli and orbital effects. 
\end{abstract} 
 
\pacs{PACS numbers: 72.15.Gd, 72.15.Eb, 72.80.Le} 
 
 
Low dimensional electronic systems characterized by a 
quasi-one-dimensional (Q1D) Fermi surface tend to form either a 
charge-density wave (CDW) or a spin-density wave (SDW) ground 
state at low temperatures as a consequence of one-dimensional 
instabilities\cite{genref,schlen}. High magnetic fields have proved to be 
useful to investigate, and even manipulate these ground states, since 
the effects are quite different for the CDW and the SDW cases. 
The Zeeman (Pauli) energy is expected to suppress  a CDW 
state because a CDW couples only bands with the same spin. In a 
magnetic field it is not possible to have the same nesting wave 
vector {\bf Q} for both spin-up and spin-down bands (see 
\cite{spuds}). In analogy with the Pauli effect in 
superconductors\cite{chan}, the Zeeman energy, 
$\mu_{\rm B}^{2}\rho(E_{\rm F})B^{2}$, (where $\rho(E_{\rm F})$ is the density 
of states at the Fermi level) competes with the CDW condensation 
energy, $-\rho(E_{\rm F})\Delta(0)^{2}$. The transition 
temperature is expected to decrease with increasing field, and 
above a certain threshold field ($ \simeq \Delta(0)/\mu_{\rm B}$) a 
uniform CDW 
is no longer energetically favorable. Consequently, a CDW may be 
suppressed by high magnetic fields. In contrast, for a SDW system,
the nesting property is not affected by the Zeeman term because a SDW 
couples spin-up with spin-down states. The nesting 
condition is actually improved by high magnetic fields due to the magnetic 
field induced one-dimensionalization of the Q1D electronic orbits. Thus 
for an imperfectly nested Fermi surface, the 
SDW transition temperature can actually increase with increasing 
magnetic field\cite{gor,ask}. The role of orbital effects on SDW 
systems has been well established in the Q1D organic Bechgaard 
salts\cite{ish}. 

By using a simple BCS relation, we can obtain a rough estimate for the 
critical field necessary to suppress a uniform CDW: $B_{c}=1.765 k_{B}/ 
\mu_{\rm B}  T_{c}$, where $k_{B}$ is the  Boltzmann constant, $\mu_{\rm B}$ is the 
Bohr magneton, and $T_{c}$ is the transition temperature to the DW state. 
However, the relatively high transition temperatures ($\geq$ 30 K) 
of most CDW systems, like for example, the molybdenum bronzes\cite{schlen} implies the need for
very high magnetic fields, of the order of 100 tesla or more, in order to suppress 
the CDW ground state via the Zeeman energy. This limitation has 
prevented the observation of this field-induced suppression.  
In this work, we argue that the $\alpha$-(ET)$_{2}$MHg(SCN)$_{4}$ (where M = K, Tl and Rb) 
organic conductors may be the {\em first} compounds whose ground state is driven
towards new DW states under the influence of {\em both} Pauli and orbital 
effects in available fields.

The band structure calculations\cite{mori} of 
$\alpha$-(ET)$_{2}$MHg(SCN)$_{4}$ 
indicate the presence of both closed Q2D and 
open Q1D orbits at the Fermi energy $E_{\rm F}$.
It is generally accepted that these systems undergo a phase transition 
from a metallic phase to a low temperature DW 
state\cite{sas1,sas2,prat} at a transition temperature, $T_{\rm DW}$, 
between 8 and 12 K. The onset of this second order transition at 
$T_{\rm DW}$ \cite{kart1}, 
is known to decrease with increasing 
field as would be expected for a CDW transition\cite{bisk}. Also, 
below $T_{\rm DW}$ and at intermediate magnetic fields (between 22 and 
37 tesla) there are profound changes in the magnetoresistance which are
indicative of a first order phase transition in the electronic 
structure at the so-called ``kink transition field", $B_{\rm K}$. This critical field clearly 
indicates that a magnetic field has a profound effect on the ground 
state of these compounds. Above 
 $B_{\rm K}$, $T_{\rm DW}$ remains finite ($\sim$ 2 K) 
\cite{bisk,kart2,kun} up to fields as high as 45 tesla\cite{hybrid}. 
After nearly a decade, the identity 
of the low temperature ground state remains a contemporary issue, 
with conflicting  evidence supporting both CDW and SDW scenarios\cite{bro}.  
There is published experimental data which, at
first glance, seems to support a SDW-like ground state: The muon
spin relaxation ($\mu$SR) rate \cite{prat} changes below $T_{\rm DW}$ while
the magnetic susceptibility is found to be anisotropic below the
same temperature \cite{sas2}. Nevertheless, no line broadening or line
splitting is observed either on the nuclear magnetic resonance
\cite{kun} nor on the electron spin resonance \cite{tsu} spectrum below
$T_{\rm DW}$. The existence of 2-D closed
orbits, clearly seen in de Haas van Alphen measurements \cite{uji}, can 
generate Landau diamagnetism which could be responsible for the
anisotropy in the magnetic susceptibility. Thus the anisotropy alone 
cannot be taken as a definitive proof for a SDW ground
state. On the other hand, no X-ray or neutron diffraction data 
that could support the existence of either a CDW or SDW 
superstructure, have thus far
been published. Clearly, there is a lack of compelling experimental
evidence providing unambiguous support for either of the two DW
ground state scenarios.

In this letter we study the angular dependence of the magnetoresistance 
and magnetization of the $\alpha$-(ET)$_{2}$MHg(SCN)$_{4}$ system. Our study reveals
new features, in particular a new magnetic-field-induced electronic
phase transition, which appears only when the angle $\theta$, defined 
as the angle between the magnetic field and the ${\bf b}^\star$ axis,
satisfies the condition $\theta = \theta_{\rm c} 
\geq 45^\circ$. Furthermore, the kink field, $B_{\rm K}$, displays a non-trivial 
angular dependence for $\theta \geq \theta_{\rm c} $.
We therefore propose a $B-\theta$ phase diagram and argue that it 
appears to be 
well explained by present theoretical models describing the behavior of a CDW 
under high 
magnetic fields with competing Pauli and orbital effects \cite{bje,zan}. 

Single crystals of $\alpha$-(ET)$_{2}$MHg(SCN)$_{4}$ (M is K, Tl, 
or Rb) were grown using conventional electrocrystallization 
techniques \cite{ish}. Transport measurements were made using four-terminal 
methods with currents ranging from 1 $\mu$A up to 10 $\mu$A applied 
perpendicular to the conducting layers (along the ${\bf 
b}^{\star}$ axis). Meanwhile, the magnetization measurements were performed using a  
phosphor-bronze cantilever magnetometer. Various configurations of cryostats, magnets, 
and rotating inserts 
available at the National High Magnetic Field Laboratory in both 
Tallahassee and Los Alamos were used in this investigation. 
 
The magnetoresistance, $R(B)$, for 
$\alpha$-(ET)$_{2}$TlHg(SCN)$_{4}$ as a function of tilted 
magnetic field, $B$, at $T \simeq 40$ mK is plotted in Fig. 1.
Figures 1 (a) and (b) show the up and down field sweeps, 
respectively. At small angles, the familiar behavior 
of the magnetoresistance as a function of field strength is observed. This includes a rapid rise 
in resistance which reaches a maximum around 15 tesla, followed by a 
drop in resistance which terminates at $B_{\rm K}$, near 27 tesla (up-sweep) 
or 24 tesla (down-sweep). $B_{\rm K}$ (indicated 
in the figure by a dashed line), is hysteretic, and is 
characteristic of a magnetic field-induced first order change in electronic 
structure. It is easily identifiable because the amplitude and wave form of the 
Shubnikov de Haas (SdH) oscillations change abruptly at 
this point. For large angles, $B_{\rm K}$ shifts to higher fields and an additional 
hysteretic structure (hereafter termed $B_{\rm c}$ 
and indicated in the figure by a dotted line) begins to appear.
Notably, $B_{\rm c}$ shifts to lower fields with increasing angle. We argue 
below that both $B_{\rm K}$ and $B_{\rm c}$ are connected with first-order transitions between 
sub-phases of the density wave ground state. In retrospect, evidence 
of $B_{\rm c}$ has been observed before, but was mislabeled as 
$B_{\rm K}$
\cite{osa}.
 
To further establish the universal character of these 
sub-phases, we provide similar results for the 
$\alpha$-(ET)$_{2}$KHg(SCN)$_{4}$ compound. Figure 2 (a) 
plots $R(B)$ as a function of $B$ (for increasing field 
sweeps) at $T \simeq$ 50 mK for several values 
of $\theta$ ($\theta$ is indicated in the figure) for a single crystal of 
$\alpha$-(ET)$_{2}$KHg(SCN)$_{4}$.
The values of $B_{\rm c}$ and $B_{\rm K}$ are 
indicated by dotted arrows and a dashed line, respectively. Both 
fields display a strong angle dependence, which is qualitatively 
similar to that discussed in Fig. 1. Notice that at $\theta = 
86^{\circ}$, $B_{\rm K}$ is outside the accessible field range. 
The behavior of  
$B_{\rm c}$  and  $B_{\rm K}$ is reproducible and observed in multiple samples. In Fig. 2 (b) 
the behavior of $B_{\rm c}$ is shown on an amplified scale for yet 
another crystal, at $T=35$ mK and for values of 
$\theta$ between $63^{\circ }$ and $90^{\circ}$. 
For the $\alpha$-(ET)$_{2}$KHg(SCN)$_{4}$ compound 
and for $\theta $ close to 60$\circ$,the field position of $B_K$ is ambiguous.
This will be the subject of future efforts.

The thermodynamic nature of $B_{\rm K}$ and $B_{\rm c}$, as transitions between 
sub-phases, was verified by magnetization measurements made on a third
sample of $\alpha$-(ET)$_{2}$KHg(SCN)$_{4}$. 
Figure 3 shows the magnetization, $M$, as a function of $B$ 
at $T = 0.5$ K for several values of $\theta$. As previously seen in 
Figs. 1 and 2(a), $B_K$ (indicated by a dashed line) moves towards 
higher fields as $\theta$ increases above $\sim 
40^{\circ}$. For fields between 12 and 20 tesla, we observe further
structure which is indicated by vertical dotted arrows
and agrees with values of $B_{\rm c}$ observed in Fig. 2. For $\theta=67^{\circ}$, both field up 
(solid line) and field down (dotted line) sweeps are included to show the 
hysteretic  behavior of both $B_{\rm K}$ and $B_{\rm c}$. Although no pronounced
discontinuities are observed in $M(B)$, the hysteretic behavior points
towards a {\em first} order phase transitions at both critical 
fields. Furthermore, the magnetization reveals additional fine structure at  
$B_{\rm c}^\prime$ and may indicate the existence of another sub-phase. 
As in Fig. 2 (a), $B_K$ can not be easily determined for $\theta$ near to 60$\circ$.

In Fig. 4, the angular dependence of both $B_{\rm c} 
(\theta)$ and $B_{\rm K}(\theta)$ are plotted for $\alpha$-(ET)$_{2}$TlHg(SCN)$_{4}$
(triangles) and for $\alpha$-(ET)$_{2}$KHg(SCN)$_{4}$ (circles).
The figure also includes $B_{\rm K}(\theta)$ obtained for an 
$\alpha$-(ET)$_{2}$RbHg(SCN)$_{4}$
single crystal (squares) at $T = 3.0$ K for fields up to 50 
tesla \cite{rbpaper}. To enable a comparison between all the three salts, we 
have normalized $B_{\rm c} (\theta)$ as well as $B_{\rm K} (\theta)$ with respect 
to the compound dependent $B_{\rm K} (\theta = 0)$. The result is a $B - 
\theta$ phase diagram containing three distinct regions. 
The hysteretic phase transition at $B_{\rm K}(\theta)$, 
indicated by a solid line, is identified as a {\em first} order 
phase transition from the zero field ground state (region 
I) to a distinct high field phase (region II). For angles larger than
$\sim 45^{\circ}$ a new phase (region III) emerges between regions I and II.
The hysteresis in Figs. 1 and 3 associated with $B_c$, indicates that the transition between regions 
II and III is also {\em first} order. The 
field dependence of $B_{\rm K}$ is very different from that of 
$B_{\rm c}$. $B_{\rm K}$ is cusp like near $\theta=90^{\circ}$.

Recently, the magnetic field dependence of a Q1D 
system with a CDW ground state was studied theoretically 
\cite{bje,zan} using a mean field approach. 
In this theory, both CDW and SDW correlations were 
included in an anisotropic 2D Hamiltonian and
studied in the random phase approximation. An important parameter 
of the theory is $\eta\equiv 
q_{0}/q_{p}=ebv_{\rm F}\cos\theta/\mu_{\rm B}$, defined as the ratio between 
the orbital and Pauli contributions to the nesting vector {\bf Q}. 
The predictions of this model strongly resemble the experimentally 
determined
phase diagram of $\alpha$-(ET)$_{2}$MHg(SCN)$_{4}$ , where M = 
K, Tl, or Rb. In particular, the theory predicts (using the author's notation) 
that: 1) below a second order transition temperature, $T_{{\rm c}0}$, the 
ground state is a uniform charge density wave $CDW_0$; 2) above a critical 
field there is a first order transition\cite{bje,zan} at $B_{{\rm c}x}$ 
to a high field state CDW$_x$ which is {\em a hybrid of charge and 
spin density wave} states; 3) between CDW$_0$ and CDW$_x$ a new 
phase CDW$_y$ is stabilized which is dependent on $\theta$ through 
$B_{{\rm c}y}\approx B_{\rm c}^{0}\sqrt{1+0.088\eta^{2}}$; and 4) all 
sub-phase transitions are first order. (CDW$_y$ is expected {\em 
not} to have SDW character). We find a close match of the above theory to 
the experimental phase diagram of $\alpha$-(ET)$_{2}$MHg(SCN)$_{4}$, if we  
assign $T_{\rm DW}$ to $T_{{\rm c}0}$ and $B_{\rm K}$ to $B_{{\rm c}x}$, 
implying that region 
I corresponds to the CDW$_{0}$ state and region II to the high field
CDW$_{x}$ phase. In effect, according to 
Ref.\cite{bje}, the Pauli effect should suppress the critical 
temperature, $T_{\rm DW}$, from the metallic phase towards region I 
(CDW${_0}$ state) in proportion to the square of the magnetic 
field \cite{spuds,bisk,bje,zan}. This is evident in the $T-B$ phase 
diagram of the $\alpha$-(ET)$_{2}$KHg(SCN)$_{4}$
compound as shown in the inset of Fig. 4. Here,
$T_{\rm DW}$ (normalized with respect to its zero field value) is 
plotted as a function of $B/B_{\rm K}$ for M = K at $\theta = 0^{\circ}$ taken from Ref. 
\cite{sas1}. In the same plot, additional data points 
are included from the onset of an abrupt change in slope of $R(B)$ as 
a function of $T$ for $B/B_{\rm K} \geq 1.1$\cite{hybrid}. Solid 
triangles indicate the position of $B_{\rm K}$ in this diagram. As has been 
previously pointed out \cite{spuds,bje,zan}, the theoretical $T-B$ phase 
diagram is remarkably similar to the diagram shown in the inset of Fig. 
4. To further strengthen the correlation between theory and experiment, we note that 
the above expression for $B_{{\rm c}y}$ may be fitted to the 
data for $B_{\rm c}$ from M = K (see the dashed line in Fig. 4) with the parameters 
$B_{\rm c}^{0}\simeq 10.4$ tesla and a Fermi velocity $v_{\rm F}=1.8\times 
10^{5}$ m/s. This is close to the value from band structure 
calculations\cite{mori} and implies well (but {\em not} perfectly) nested Q1D Fermi 
sheets for the $\alpha$-(ET)$_{2}$MHg(SCN)$_{4}$ salts (where M = K, 
Tl, or Rb). 
 
In summary, we have closely examined the angular-dependent 
magnetoresistance and magnetization in the low temperature DW
ground state of $\alpha$-(ET)$_{2}$MHg(SCN)$_{4}$ (where M = 
K, Tl, or Rb). We find that the material exhibits at least three low 
temperature electronic sub-phases, which are separated by first order phase 
boundaries. We argue that for low fields and tilted angles, the 
ground state is well represented by a CDW
description, (albeit that no direct evidence for a CDW as opposed to a 
SDW presently exists). For $\theta\geq\theta_{\rm c}\simeq 45^{\circ}$ 
we identify a new structure, $B_c$, seen in the angular dependence of 
both magnetoresistance and magnetization, as a field-induced phase 
transition governed by the competition between orbital and Pauli 
effects. The appearance of this new phase at $\theta_{\rm c}$ 
displaces $B_{\rm K}$ (given by the Pauli 
limit) towards higher values. The $B-T$ and $B-\theta$ phase diagrams are well described by the 
available models for charge-density waves in high magnetic fields. This study both 
supports theoretical predictions of the complex behavior of CDW in a 
magnetic field and clarifies the nature of the ground 
state in the $\alpha$-(ET)$_{2}$MHg(SCN)$_{4}$ compounds. 
 
The authors thank N. Bi\v{s}kup for helpful discussions, and 
recognize the cooperation received from S. Y. Han, Brian H. Ward, Yuri 
Sushko, and the staff of NHMFL-LANL. In addition, we acknowledge support from NSF-DMR 
95-10427 and 99-71474 (JSB). One of us (LB) is grateful to the 
NHMFL for sabbatical leave support. The NHMFL is supported through 
a cooperative agreement between the State of Florida and the 
NSF through NSF-DMR-95-27035.

\begin{figure}[htbp] 
\caption{(a) Magnetoresistance, $R(B)$, of a 
$\alpha$-(ET)$_{2}$TlHg(SCN)$_{4}$ single crystal as a function 
of magnetic field $B$, at $T=40$ mK, for increasing field sweeps 
at several angles $\theta$ between $B$ and ${\bf b}^*$ ($\theta$ 
is indicated in the figure). (b) Same as in (a) but for decreasing 
field sweeps. Dashed line indicates $B_{\rm K}$ while the dotted line 
indicates $B_{\rm c}$. In both figures curves are vertically displaced 
for clarity.} 
\end{figure}

\begin{figure}[htbp] 
\caption{(a) $R(B)$ for a $\alpha$-(ET)$_{2}$KHg(SCN)$_{4}$ single 
crystal as a function of $B$ at $T=50$ mK for increasing field 
sweeps and several values of $\theta$ (indicated in the figure). 
$B_{\rm K}$ is indicated by a dashed line while dotted arrows indicate 
$B_{\rm c}$. (b) $R(B)$ as a function of $B$, on an amplified scale, for 
a second $\alpha$-(ET)$_{2}$KHg(SCN)$_{4}$ single crystal at $T=35$ mK
and for different values of $\theta$ as indicated. The line indicates 
$B_{\rm c}$. 
In both figures, curves are vertically 
displaced for clarity while in (a) all curves for $\theta \geq 
52^{\circ}$ are multiplied by a factor of 5.} 
\end{figure}

\begin{figure}[htbp] 
\caption{Magnetization, $M$, of a 
$\alpha$-(ET)$_{2}$KHg(SCN)$_{4}$ single crystal as a function of 
$B$ at $T \simeq$ 500 mK and for four values of $\theta$. All 
curves are displaced vertically while the curves at $\theta =$ 
60$^{\circ}$ and 67$^{\circ}$ are multiplied by a factor of 5. 
$B_{\rm K}$ is indicated by a dashed line while $B_{\rm c}$ is indicated by 
both dotted line and dotted vertical arrows. Solid arrows indicate 
the place of $B_{\rm K}$ for $\theta = 60^{\circ}$. All solid lines are for increasing 
field sweeps.
The dotted line at $\theta = 67^{\circ}$ indicates a decreasing field sweep. 
$B_{\rm c}^\prime$, also indicated by a dashed
line, suggests an additional phase transition.} 
\end{figure}

\begin{figure}[htbp] 
\caption{(a) $B_{\rm c}(\theta )$ as well as $B_{\rm K}(\theta )$, both 
normalized respect $B_{\rm K}(\theta=0 )$, for each sample shown in Figs. 
1 and 2. Solid and opened triangles are $B_{\rm c}(\theta )$ and 
$B_{\rm K}(\theta )$, respectively, obtained from Fig. 1. Similarly, 
solid and opened circles were obtained from Fig. 2 and other $\alpha$-(ET)$_{2}$KHg(SCN)$_{4}$ 
samples, while squares correspond to $B_{\rm K}$ measured in a 
$\alpha$-(ET)$_{2}$RbHg(SCN)$_{4}$ sample at $T=3.0$ K. The resulting $B-\theta$ phase diagram is 
composed of three regions. Solid lines are guides to the eyes and 
suggest first order phase transitions. The dashed line (also indicates first order) is a fit to 
the expression for $B_{{\rm c}y}$, see the text. Inset: $T_{\rm DW}$ from Ref. [9] 
normalized with respect to $T_{\rm DW}$ at zero field, as a function of 
$B/B_{\rm K}$ for M = K (circles). 
We added new points for $B/B_{\rm K} \geq 1.1$ as well as the position 
of $B_{\rm K}$ in this phase diagram (solid triangles).} 
\end{figure}


\begin{references} 
 
 
\bibitem{genref} V. J. Emery, in {\em Highly Conducting One Dimensional 
Solids}, edited by J. T. Devreese, R. P. Evrard, and V. E. Van 
Doren (Plenum, New York, 1979), p. 247; J. Solyom, Adv. Phys. {\bf 
28}, 201 (1979); {\em Electronic Properties of Inorganic 
Quasi-One-Dimensional compounds}, edited by P. Monceau (Reidel, 
Dordrecht, 1985). 
 
\bibitem{schlen} C. Schlenker, in {\em Low Dimensional Electronic 
Properties of Molybdenum Bronzes and Oxides}, (Kluwer, New York, 
1989). 

\bibitem{spuds} R. H. McKenzie, cond-mat/9706235 (1997). 
 
\bibitem{chan} B. S. Chandrasekhar, Appl. Phys. Lett. 1, {\bf 7} (1962); A. 
M. Clogston, Phys. Rev. Lett {\bf 9}, 266 (1962). 
 
\bibitem{gor} L. P. Gor'kov and A. G. Lebed, J. Phys. Lett. (Paris) {\bf 
45}, L433 (1984). 
 
\bibitem{ask} A. Audouard and S. Askenazy, Phys. Rev. B {\bf 52}, R700 (1995); 
 A. Bjelis and K. Maki, Phys. Rev B {\bf 44}, 6799 (1991). 
 
\bibitem{ish} T. Ishiguro, J. Yamaji, and G. Saito, in {\em Organic 
Superconductors}, (Springer-Verlag, Berlin, Heildeberg, New 
York,1998); J. Wosnitza, {\em Fermi Surfaces of Low-Dimensional 
Organic Metals and Superconductors}, (Springer-Verlag, Berlin, 
Heildeberg, New York,1996). 
 
\bibitem{mori} H. Mori et {\em al.}, Bull. Chem. Soc. Jpn. {\bf 63}, 2183 
(1990); L. Ducasse and A. Frisch, Solid Sate Comm. {\bf 91}, 201 
(1994); R. Rousseau et al. J. Phys (France) I {\bf 6}, 1527 
(1996). 
 
\bibitem{sas1} T. Sasaki {\em et al.}, Solid State Comm. {\bf 75}, 93 
(1990); T. Sasaki, S. Endo, and N. Toyota, Phys. Rev. B {\bf 48}, 
1928 (1993); P. F. Henning {\em et al.}, Solid State Comm. {\bf 
95}, 691 (1995); M. K\"{o}ppen {\em et al.}, Synth. Met. {\bf 86}, 
2057 (1997). 
 
\bibitem{sas2} T. Sasaki, H. Sato, and N. Toyota, Synth Met. {\bf 41-43}, 
2211 (1991); P. Christ et al., Synth. Met. {\bf 86}, 2057 (1997). 
 
\bibitem{prat} F. L. Pratt {\em et al.}, Phys. Rev. Lett. {\bf 74}, 3892 
(1995). 
 
\bibitem{kart1} P. Henning {\em et al.}, Solid State Commun. {\bf 95}, 691 (1995).
 
\bibitem{bisk} N. Bi\v{s}kup, J. A. A. J. Perenboom, 
J. S. Brooks and J. S. Qualls, Solid State Comm. {\bf 107}, 503 
(1998); M. V. Kartsovnik {\em et al.}, Synth Met. {\bf 86}, 1933 
(1997); T. Sasaki, A. G. Lebed, T. Fukase, and N. Toyota, Phys. 
Rev. B {\bf 54}, 12969 (1996). 
 
\bibitem{kart2} M. V. Kartsovnik {\em et al.}, Synth. Met. {\bf 86}, 1933 (1997).
 
\bibitem{kun} P. L. Kuhns {\em et al.}, Solid State Commun. {\bf 109}, 637 
(1999); T. Takahashi {\em et al.}, Synth. Met. {\bf 55-57}, 2513 
(1993); K. Kanoda {\em et al.}, Synth. Met. {\bf 70}, 973 (1995); 
K. Miyagawa, A. Kawamoto, and K. Kanoda, Synth. Met. {\bf 86}, 
1987 (1997). 

\bibitem{hybrid} J. S. Brooks {\em et al.}, to be published. 
 
\bibitem{bro} J. S. Brooks {\em et al.}, in Proceedings of Physical Phenomena 
at High Magnetic Fields II, edited by Z. Fizk, L. Gor'kov, D. 
Meltzer and R. Schrieffer, (World Scientific, Singapore, 1996), p. 
249. 

\bibitem{tsu} R. Tsuchyia {\em et al.}, Synth. Met. {\bf 70}, 965 (1995).

\bibitem{uji} S. Uji {\em et al.}, Solid State Comm. {\bf 12}, 825 (1996).
 
\bibitem{bje} A. Bjeli\v{s}, D. Zanchi, and G. Montambaux, Journal de 
Physique IV 9:(P10), 203 (1999); A. Bjeli\v{s}, D. Zanchi, and G. 
Montambaux, Cond-mat/9909303 21 Sep 1999. 
 
\bibitem{zan} D. Zanchi, A. Bjeli\v{s}, and G. Montambaux, Phys. Rev. B 
{\bf 53}, 1240 (1996). 
 
\bibitem{osa} T. Osada, R. Yagi, A. Kawasumi, and S. Kagoshima, Synth 
Met. {\bf 41-43}, 2171 (1991).
 
\bibitem{rbpaper} J. S. Qualls {\em et al.}, to be published. 

\end{references}
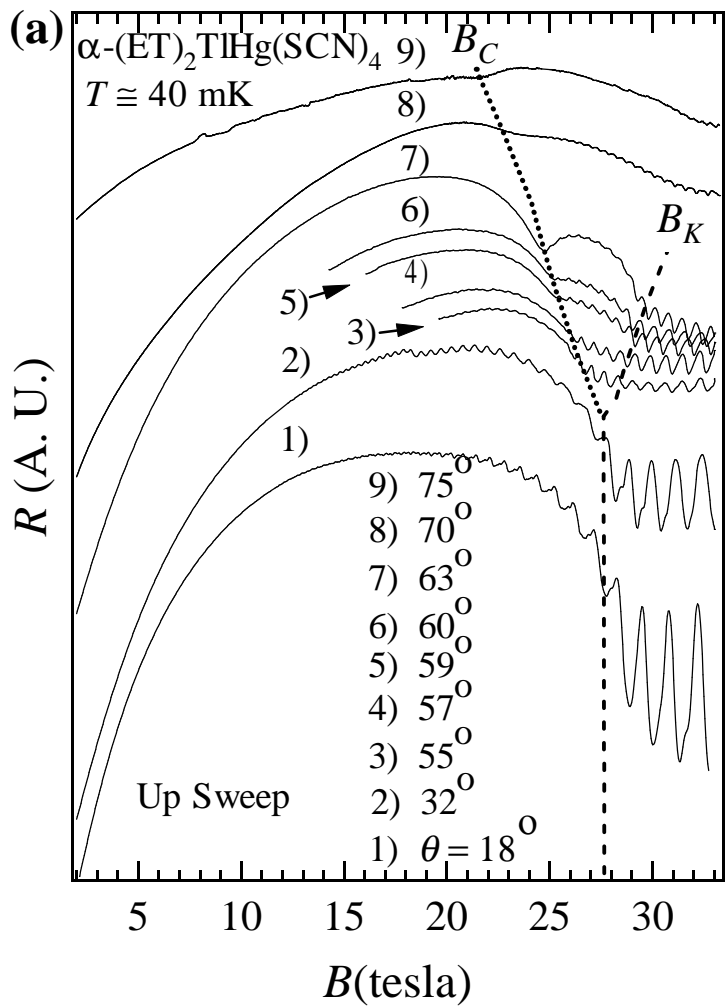
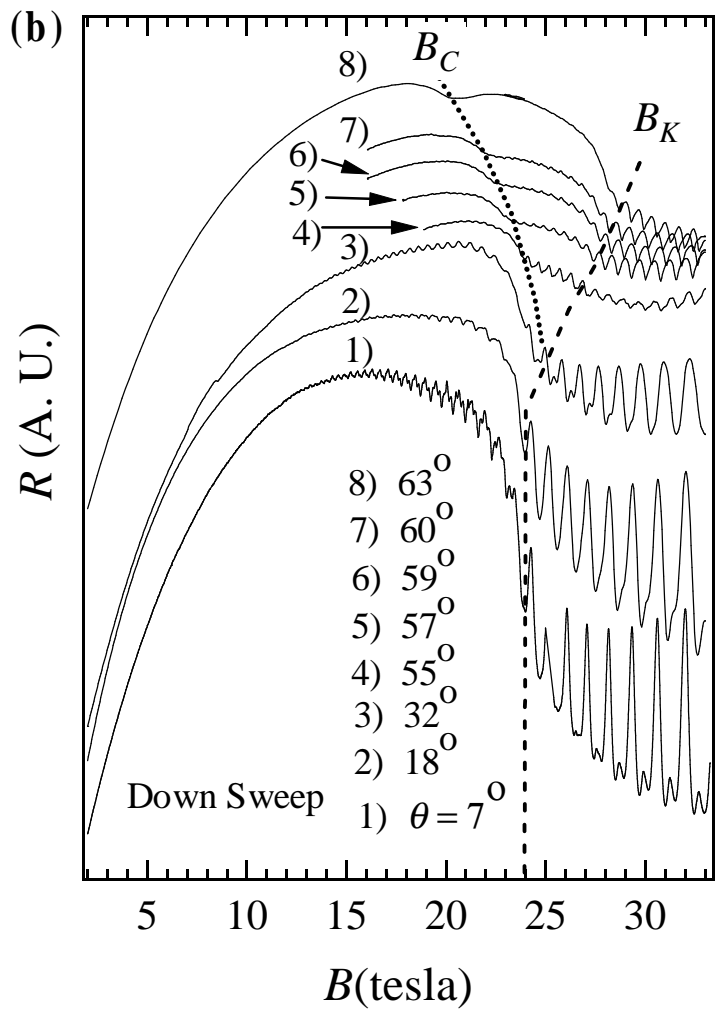
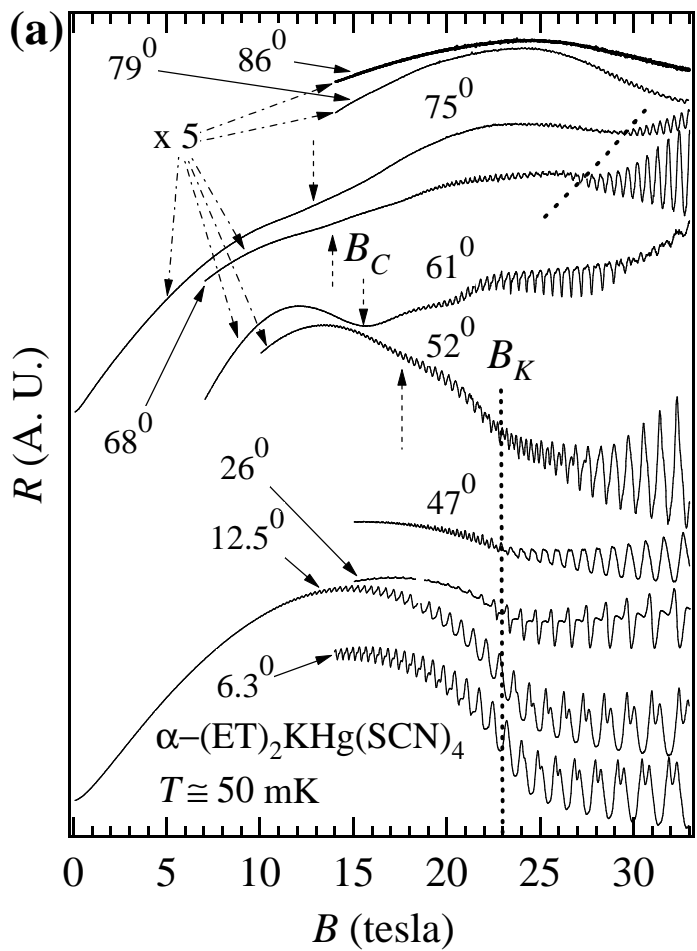
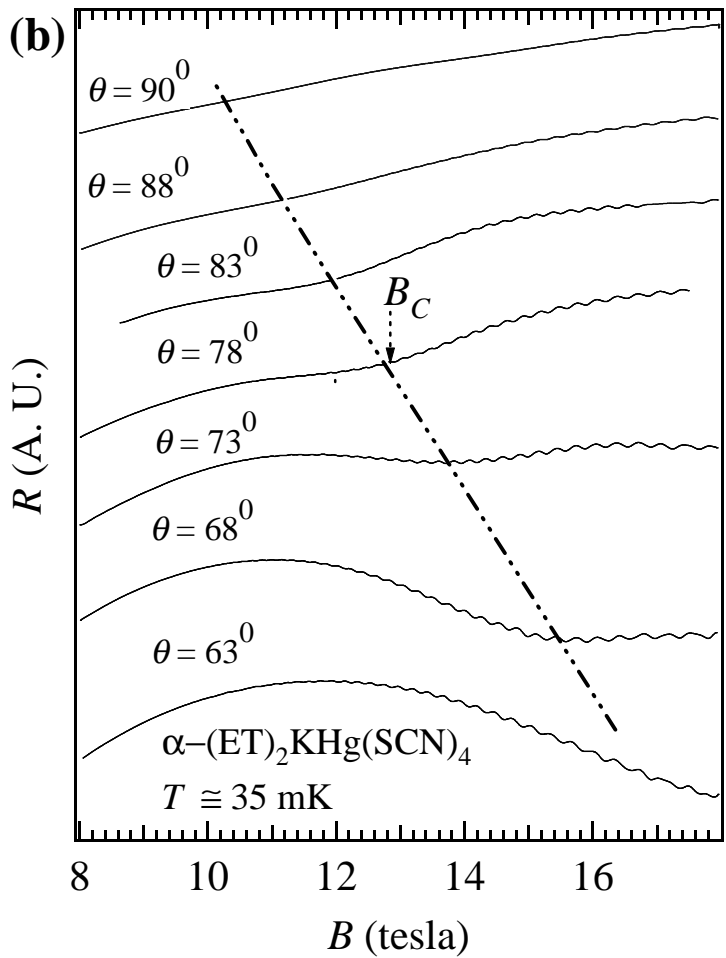
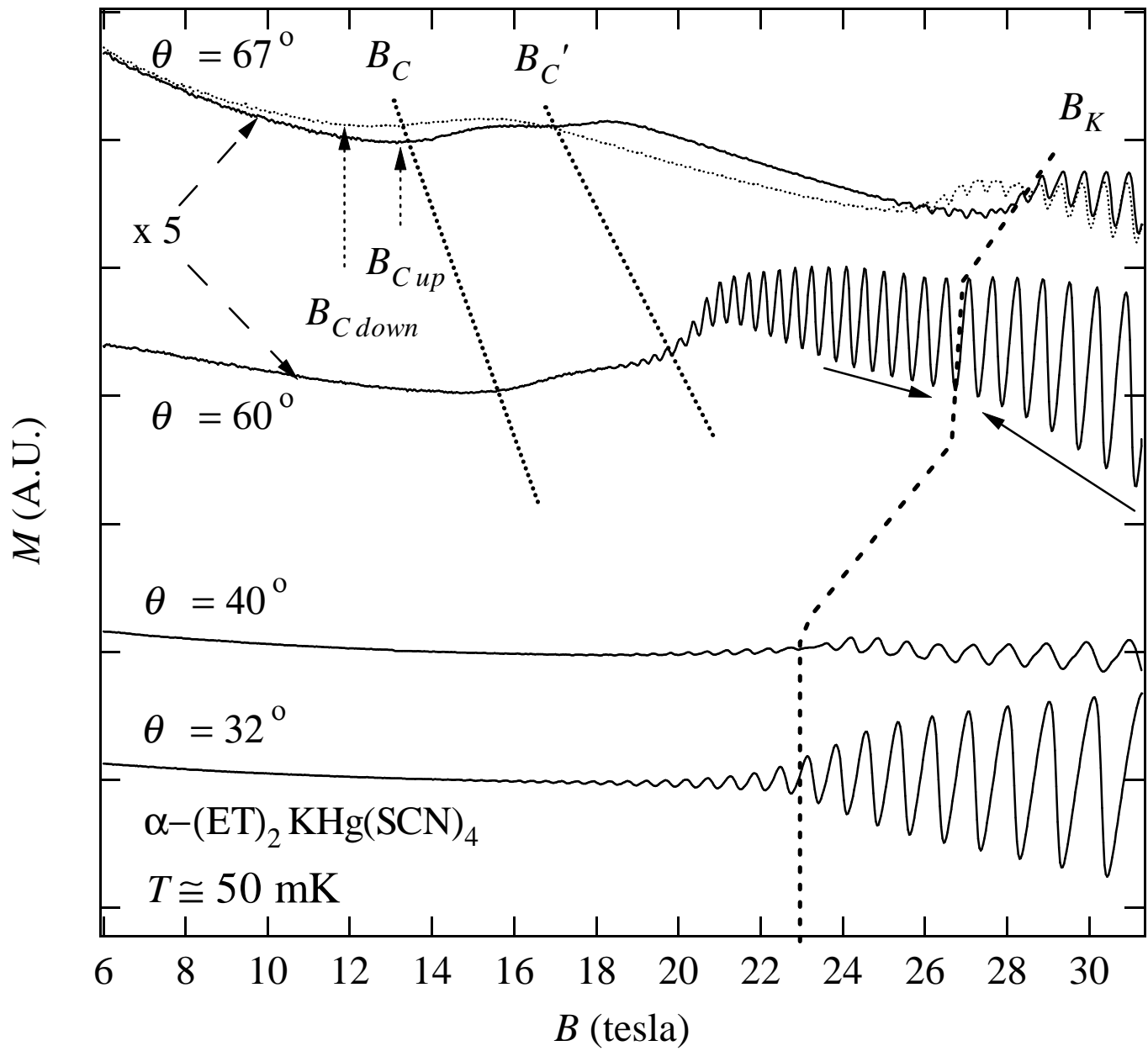
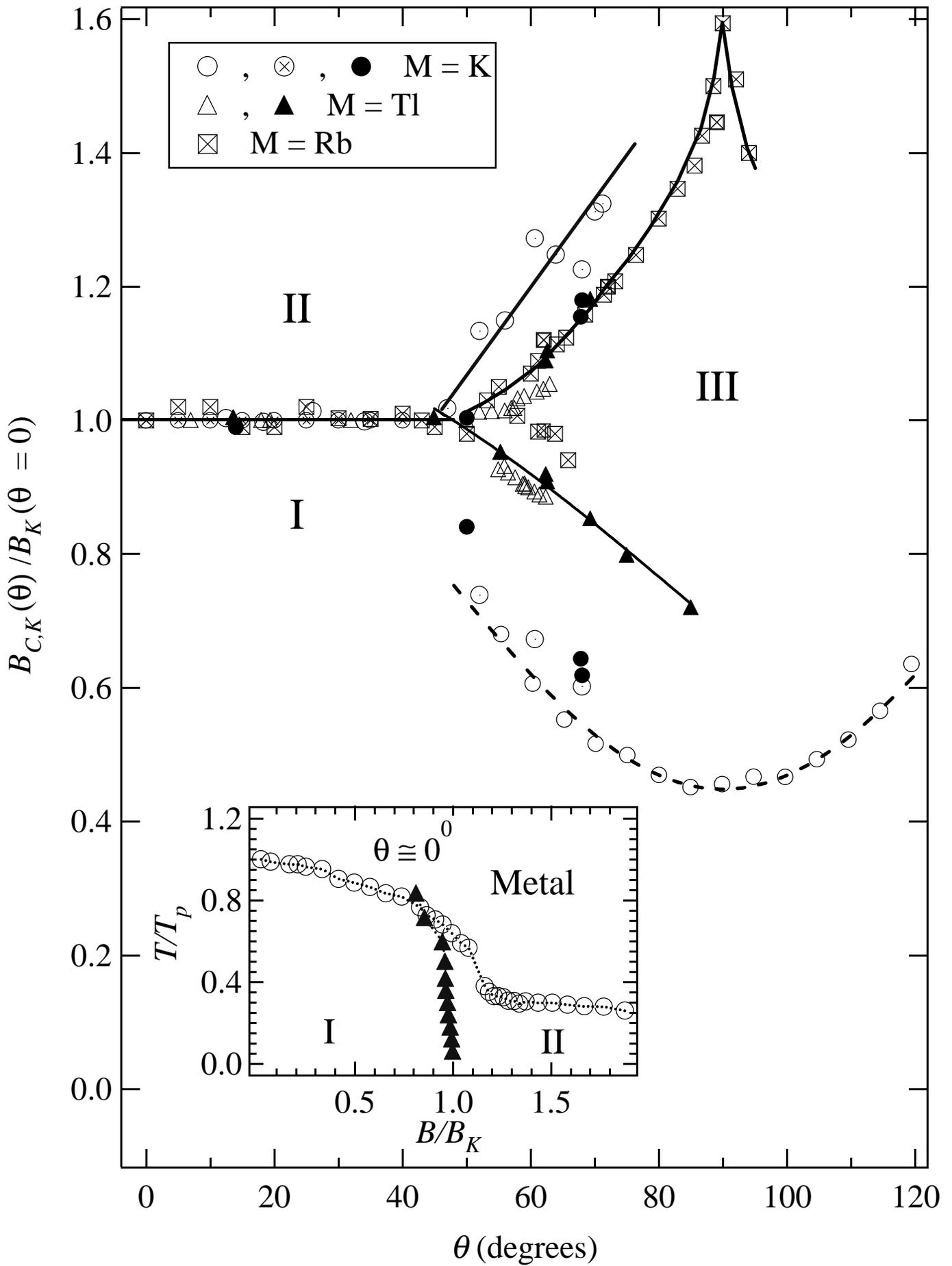
\end{document}